\documentclass[prl,twocolumn,amsmath,amssymb,showpacs]{revtex4}

\usepackage{epsfig}
\usepackage{graphicx}

\begin{document}

\title{Choice of measurement as signal}

\author{Amir Kalev}
\affiliation{Centre for Quantum Technologies, National University of Singapore, 3 Science Drive 2, 117543, Singapore\\Center for Quantum Information and Control, University of New Mexico, Albuquerque, NM 87131-0001, USA\footnote{Current address}}
\author{Ady Mann}
\author{Michael Revzen}
\affiliation {Department of Physics, Technion - Israel Institute of Technology, Haifa 32000, Israel}

\date{\today}

\begin{abstract}

In classical mechanics, performing a measurement without reading the measurement outcome is
equivalent to not exploiting the measurement at all. A non-selective measurement in the
classical realm carries no information. Here we show that the situation is remarkably
different when quantum mechanical systems are concerned. A non-selective measurement on one
part of a maximally entangled pair can allow communication between two parties. In the
proposed protocol, the signal is encoded in the {\it choice} of the measurement basis of
one of the communicating parties, while the outcomes of the measurement are irrelevant for
the communication and therefore may be discarded. Different choices for the (non-selective)
measurement correspond to different signals. Implication to the study of measurements
in quantum mechanics is considered. The scheme is studied in a Hilbert space of prime
dimension.
\end{abstract}

\pacs{03.65.Ta;03.67.Hk}

\maketitle

{\it Introduction---} An important distinction between classical and quantum measurement is that the latter
implies an inevitable disturbance to the measured system. In the present work we show that
this disturbance is trackable to the extent that it may be used for communication. Thus we
study non-selective measurements where the outcomes are not recorded. Such measurements
within the classical theory do not carry information and hence cannot be used for
communication \cite{schwinger01,diosi11}. Here we show how non-selective measurements on
one part of a quantum system of a maximally entangled pair can be used to encode and
eventually communicate information. In the proposed protocol the {\it basis,
i.e. the choice}, of the (non-selective) measurement is the signal. The outcomes of the
measurement are totally irrelevant.\\
 The trackable choice of measurement (rather than its outcome) analysis allows a novel
interpretation of quantum measurements. Thus a quantum state does, in general,
\cite{kochen, bell, mermin} imply contextual values for measurable dynamical variables.
Hence it is attractive to interpret our result as suggesting that quantum measurement is
built up of two stages. The first stage, to be associated with the non-selective
measurement, elevates a particular set of dynamical variables (those labelling the basis in
our case) to reality (i.e. having a prescribed value). The state after this stage is, in
general, mixed. The second stage involves the determination of the value of the dynamical
variable. This stage has clear classical attributes.

Confining our study at the moment to a Hilbert space of odd prime dimension, $d$, we
consider as alternative choices for measurements the alternative mutual unbiased bases
(MUB). For prime dimension there are  $d+1$ MUB \cite{ivanovich,
vourdas97,wootters89,tal,gibbons04,vourdas04,durt10}. A possible set of $d{+}1$ MUB can be
defined as follows. The first basis is the computational basis
$\{|n\rangle\}_{n{=}0}^{d{-}1}$, composed of the $d$ orthonormal eigenstates of the
generalized Pauli operator $\hat{Z}$, $\hat{Z}|n\rangle=\omega^n|n\rangle, |n{+}d\rangle=
|n\rangle, \omega= e^{i\frac{2\pi}{d}}$. The other $d$ orthonormal bases are parametrized
by $b{=}0,1,\ldots,d{-}1$. The kets that compose the $d$ remaining bases are given in terms
of the computational basis by \cite{tal}
\begin{equation}
|m;b\rangle=\frac{1}{\sqrt d}\sum_{n=0}^{d-1}|n\rangle\omega^{bn^2-2nm};\;\;b,
m=0,1,\ldots,d-1.
\end{equation}
We shall designate the computational basis by $b=\ddot{0}$, and depending on the context we may also
denote the kets of the computational basis $|m\rangle$ by $|m;\ddot{0}\rangle$. Thus, the $d{+}1$ bases
are labelled by
$b=\ddot{0},0,1,\ldots,d{-}1$.

The proposed communication protocol is described in what follows. We
assume that the two communicating parties, Alice and Bob, agree beforehand upon a code,
associating messages with the parameters, $b$, specifying the MUB. There is no classical
communication between Alice and Bob beyond this point. The protocol involves a two
$d$-level system (qudit) entangled state prepared by Alice with one qudit available to Bob
who wishes to communicate a message, $b=\ddot{0},0,1,\ldots,d{-}1$,  to Alice. To this end
Bob measures the part of the system that is available to him in the basis parametrized with
the $b$ of his message. He must complete the measurement yet may ignore (!) its outcome and
then return the qudit to Alice. This step renders values to a class of dynamical variables:
 the complete set of commuting operators $(XZ^b)^n;\;n=0,1,...d-1$. Now Alice measures the two-qudit resultant state and deduces, almost always, the basis $b$
used by Bob, hence decodes the message. The procedure is quantal in that the signal
corresponds to the {\it basis} of Bob's measurement, that is to the ``alignment'' of his
instrument, and in that the measurement outcomes are irrelevant and may be unrecorded.

{\it Choice of measurement basis as signal---} To establish a communication channel, let Alice prepare one of the following $d^3$ two-qudit maximally
entangled states \cite{rev1,rev2},
\begin{equation}\label{crs}
|c,r;s\rangle_{1,2}= \frac{1}{\sqrt
d}\sum_{n=0}^{d-1}|n\rangle_1|c-n\rangle_2\,\omega^{sn^2-2rn},
\end{equation}
with $c,r,s=0,1,\ldots,d{-}1$, and send one of the qudits, say, the one labelled by 1, to
Bob. We note that, for a given $s$ value, these states form an orthonormal, maximally
entangled, basis for the Hilbert space of the two qudits. Thus, $s$ labels the basis and
$c$ and $r$ label the $d^2$ orthonormal states within a basis. The reduced state for Bob's qudit is the completely mixed state.

\begin{figure}[t]
\centering
\includegraphics[scale=0.5]{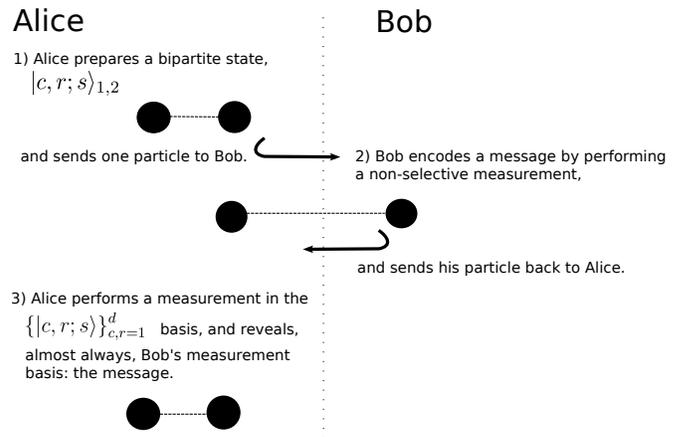}
\caption{The scheme of the communication protocol}
\label{fig:protocol}
\end{figure}

To communicate a message to Alice, Bob measures his qudit in one of the MUB labelled by $b=\ddot{0},0,1,\ldots,d{-}1$. The message is his choice of the basis used for the measurement.  Bob may or may not record the
measurement outcome. This is of no relevance to the protocol. After completing his non-selective measurement, Bob sends the qudit back to Alice. The two-qudit state is described now by
\begin{equation}\label{rho12}
\rho_{1,2}=\sum_{m=0}^{d-1}|m;b\rangle_1\langle m;b|c,r;s\rangle_{1,2}\langle
c,r;s |m;b\rangle_1\langle m;b|.
\end{equation}
We note in passing that making a non-selective measurement in basis $b$ is equivalent to performing a
random unitary transformation which is diagonal in the $b$ basis \cite{schwinger01}.
To retrieve the message, Alice now measures the two qudits in the  basis of preparation,
$\{|c',r';s\rangle_{1,2}\}_{c',r'=0}^{d-1}$ of Eq.~(\ref{crs}). The probability to obtain
an outcome which corresponds to the basis state $|c',r';s\rangle_{1,2}$ is
\begin{align}
&\langle
c',r';s|\rho_{1,2}|c',r';s\rangle_{1,2}\\\nonumber&=\frac{1}{d}\left\{
  \begin{array}{ll}
    \delta_{c,c'} & \textrm{for}\; b=\ddot{0}\\
    \delta_{(b-s)c+r,(b-s)c'+r'} & \textrm{for}\; b=0,1,2,\ldots,d-1.
  \end{array} \right.
\end{align}
The arithmetics is modulo $d$. According to the above equation,  based on the outcome of her measurement,
Alice can decode the message sent from Bob, that is, the basis of his measurement. If the
outcome corresponds to a state $|c',r';s\rangle_{1,2}$ with $c\neq c'$, Alice infers that
$b=s{+}\frac{r'{-}r}{c{-}c'}$. Since  she knows the values of $c,r,c',r'$ and $s$, she can
calculate the message $b$. If, on the other hand, $c=c'$ and $r\neq r'$ Alice infers that
$b=\ddot0$. The case of $c=c'$ and $r=r'$ is inconclusive for Alice. The inconclusive
outcome occurs with probability $1/d$. In that case the preparation state and the
detection state of the two qudits is the same and she does not gain any  information about
Bob's message. Hence the decoding table is,
\begin{equation}\label{b}
 \begin{array}{rcl}
  c\neq c' & {\rightarrow} & b=s+\frac{r-r'}{c'-c} \\
  r\neq r',\;c = c' & {\rightarrow} & b=\ddot0 \\
  r = r',\;c = c' & {\rightarrow} & {\rm inconclusive}.
 \end{array}
\end{equation}
For the even prime dimension, $d{=}2$, by plugging the imaginary unit $i$ instead of
$\omega$ in all of the above equations, one retrieves the same decoding table~(\ref{b}). The protocol is schematically drawn in Fig.~\ref{fig:protocol}.

{\it Conclusions and remarks---} To conclude, we showed how non-selective measurements in MUB on one part of an entangled pair could be used to encode information. The scheme uniquely utilizes quantum features of the system, since performing non-selective measurements on classical systems (no matter how correlated they are) cannot carry or manipulate information \cite{schwinger01,diosi11}.
Alternatively the trackability of the non-selective measurement allows a novel view of
quantum measurement. Thus we may view a quantum measurement as a two stage process. The
first, to be associated with the non-selective part, involves the promotion of a set of
dynamical variables (labelled, in our case, by the basis b) to reality (i.e. \cite{mermin}
having a definite value). After this stage, in general, the state is a mixed state. The
second stage involves the determination of the outcome among these values. This stage
allows a classical interpretation: the experiment determines a possible preassigned value.
Dealing, as we do in this work, with entangled state renders the stages separable: Bob's
measurement (unknown to Alice) consummates a first stage {\it for the two-particle system}.
The sequential measurement, by Alice, selects possible values of the two particle system
allowed by the (mixed) state that resulted from Bob's measurement.\\

In the considered protocol, by sending a qudit, Bob is able to transfer, on average, more
than $\log_2{d}$ bits of information to Alice. This is, in some respect, a form of dense
coding. We note for comparison that super-dense coding \cite{dense92} achieves $2\log_2{d}$
bits per qudit sent from Bob to Alice. However, in the super-dense coding scheme specific
unitary transformations are used for the encoding, while here non-selective measurements,
or, equivalently, random unitary transformations, are utilized. Though, in its present
form, the protocol cannot be used for secure communication, we leave it as an open question
whether one could consider variations of the present protocol that would render it suitable
for cryptography tasks. Preliminary study indicates that the proposed scheme can be generalized to encompass prime-powers dimensions. Finally, this protocol exemplifies how tasks which seem impossible by classical reasoning are realized in quantum systems.

\begin{acknowledgments}
AK would like to thank Prof. B.-G. Englert for fruitful discussions and for his insightful
comments. The Centre for Quantum Technologies is a Research Centre of Excellence funded by
the Ministry of Education and by the National Research Foundation of Singapore. This research was supported in part by NSF Grants No. PHY-1212445.
\end{acknowledgments}

\end{document}